\documentclass[11pt]{article}
\usepackage{amsmath,amssymb}
\usepackage[all]{xy}
\input diagxy

\font\fr=eufm10 scaled \magstep 1 %(caracteres goticos)
\newtheorem{teor}{Theorem}
\newtheorem{prop}{Proposition}

\newtheorem{lem}{Lemma}
\newtheorem{definition}{Definition}

\def\beq{\begin{equation}}
\def\eeq{\end{equation}}
\def\bea{\begin{eqnarray}}
\def\eea{\end{eqnarray}}
\def\beann{\begin{eqnarray*}}
\def\eeann{\end{eqnarray*}}
\def\beasn{\begin{sneqnarray}}
\def\eeasn{\end{sneqnarray}}
\def\ben{\begin{enumerate}}
\def\een{\end{enumerate}}
\def\bit{\begin{itemize}}
\def\eit{\end{itemize}}
\def\proof{ (\emph{Proof\/}) }
\newcommand{\ds}{\displaystyle}
\def\derpar#1#2{\frac{\partial{#1}}{\partial{#2}}}

\def\qed{\ifvmode\removelastskip\fi
{\unskip\nobreak\hfil\penalty50\hbox{}\nobreak\hfil
\hbox{\vrule height1.2ex width1.2ex}\parfillskip=0pt
\finalhyphendemerits=0 \par\smallskip}}
\def\vf{\mbox{\fr X}}
\def\df{{\mit\Omega}}

\def\Lden{{\cal L}}

\def\d{{\rm d}}

\def\Real{\mathbb{R}}
\def\R{\mathbb{R}}

\def\Tan{{\rm T}}

\def\inn{\mathop{i}\nolimits}
\def\Cinfty{{\rm C}^\infty}
\def\tabaddress#1{{\small\it\begin{tabular}[t]{c}#1
\\[1.2ex]\end{tabular}}}

\title{UNIFIED FORMALISM \\
  FOR NON-AUTONOMOUS MECHANICAL SYSTEMS}
\author{\sc Mar\'\i a Barbero-Li\~n\'an\thanks{{\bf e}-{\it mail}:
  mbarbero@ma4.upc.edu},
  Arturo Echeverr\'\i a-Enr\'\i quez\thanks{{\bf e}-{\it mail}:
  arturo@ma4.upc.edu},
  \\
  \tabaddress{Departamento de Matem\'atica Aplicada IV\\
  Edificio C-3, Campus Norte UPC\\
  C/ Jordi Girona 1. 08034 Barcelona. Spain}
  \\
{\sc David Mart\'\i n de Diego\thanks{{\bf e}-{\it mail}:
d.martin@imaff.cfmac.csic.es}} \\
\tabaddress{Instituto de Ciencias Matem\'aticas (CSIC-UAM-UCM-UC3M)\\
   C/ Serrano 123. 28006 Madrid. Spain}
     \\
{\sc Miguel C. Mu\~noz-Lecanda\thanks{{\bf e}-{\it mail}:
  matmcml@ma4.upc.edu}},
{\sc Narciso Rom\'an-Roy\thanks{{\bf e}-{\it mail}:
  nrr@ma4.upc.edu}},
  \\
  \tabaddress{Departamento de Matem\'atica Aplicada IV\\
   Edificio C-3, Campus Norte UPC\\
   C/ Jordi Girona 1. 08034 Barcelona. Spain}}
\begin{document}

\maketitle

\pagestyle{myheadings}

\thispagestyle{empty}

\begin{abstract}
We present a unified geometric framework for describing
both the Lagrangian and Hamiltonian formalisms of regular
and non-regular time-dependent mechanical systems, which is
based on the approach of Skinner and Rusk \cite{SR-83}. The
dynamical equations of motion and their compatibility and
consistency are carefully studied, making clear that all
the characteristics of the Lagrangian and the Hamiltonian
formalisms are recovered in this formulation. As an
example, it is studied a semidiscretization of the
nonlinear wave equation proving the applicability  of the
proposed formalism.
\end{abstract}

 % \bigskip

\bigskip

\vbox{\raggedleft AMS s.\,c.\,(2000): 37J05, 53D05, 55R10, 70H03, 70H05 }\null

\markright{\sc M. Barbero-Li\~n\'an {\it et al\/},
    \sl Skinner-Rusk formalism for non-autonomous systems}

 % \clearpage

%\tableofcontents

%\newpage

\section{Introduction}

In 1983 Skinner and Rusk introduced a representation
of the dynamics of an autonomous mechanical system which combines
the Lagrangian and Hamiltonian features \cite{SR-83}.
The aim of this formulation was to obtain a common
framework for both regular and singular dynamics,
obtaining simultaneously the Hamiltonian and Lagrangian formulations of the
dynamics. Over the years, however, Skinner and
Rusk's framework was extended in many directions. So, Cantrijn {\it
et al} \cite{CMC-2002} extended this formalism for explicit
time-dependent systems using a jet bundle language.
In \cite{GM-05} an extension of this
formalism to other kinds of more general time-dependent singular
differential equations was given. Cort\'es {\it
et al} \cite{CLMM-2002} used the Skinner and Rusk formalism to
consider vakonomic mechanics and the comparison between the
solutions of vakonomic and nonholonomic mechanics. Finally, in
\cite{ELMMR-04,LMM-2002,RRS} the Skinner-Rusk
model was developed for classical field theories.

The aim of this paper is to continue the study of the
the Skinner-Rusk formalism for time dependent mechanical
systems (Section \ref{uf}), now, carefully studying the dynamical equations of
motion and the submanifolds where they are consistent,
and showing how the Lagrangian and Hamiltonian descriptions
are recovered from this unified framework
(Sections \ref{des},\ref{devf}).

The case of field theories was independently developed in
\cite{ELMMR-04,LMM-2002}, and improves the construction
given in \cite{CMC-2002}, as it is discussed in Section
\ref{di}.

As a new application, we analyze the case of semidiscretizations of
field theories in Section \ref{semi}. These methods are designed by
numerical schemes that respect physical principles preserved by the
continuous systems, specially those described by partial
differential equations (PDEs). In this case, there are not only a
time dependence (as in ordinary differential equations) but also
posses an spatial dependence. Many integration methods, in
particular in Hamiltonian dynamics, starts by discretizing the
spatial structure (spatial truncation)  obtaining a  finite
dimensional system of ordinary differential equations (ODEs)
retaining some physical properties of the original system (see
\cite{LeRe}). For simplicity, we restrict ourselves to a particular
semidiscretization of the nonlinear wave equation
\cite{LMM-2008,OlWeWu} obtaining a unique solution of the dynamics
on the secondary constraint submanifold.

All the manifolds are real, second countable and ${\cal
C}^{\infty}$. The maps are assumed to be ${\cal
C}^{\infty}$. Sum over repeated indices is understood.

\section{Non-autonomous Lagrangian and Hamiltonian systems}
\protect\label{nads}

(See \cite{EMR-91,Ku-tdms,MS-98,Ra1,St-2005} for more
details). In the jet bundle description of non-autonomous
dynamical systems, the configuration bundle is $\pi\colon
E\to\Real$, where $E$ is a $(n+1)$-dimensional
differentiable manifold endowed with local coordinates
$(t,q^i)$, and $\Real$ has $t$ as a global coordinate. The
jet bundle of local sections of $\pi$, $J^1\pi$, is the
{\sl velocity phase space} of the system, with natural
coordinates $(t,q^i,v^i)$, adapted to the bundle $\pi\colon
E\to\Real$, and natural projections are
$$
\pi^1 \colon J^1\pi \to E \quad ,\quad \bar\pi^1 \colon J^1\pi \to
\Real \, .
$$
(If $E\equiv \Real\times Q$, where $Q$ is a
$n$-dimensional differentiable manifold, then
$J^1\pi\simeq \Real\times\Tan Q$).

A Lagrangian density $\Lden\in\df^1(J^1\pi)$ is a
$\bar\pi^1$-semibasic $1$-form on $J^1\pi$, and it is usually
written as $\Lden = L\,\d t$, where $L\in\Cinfty (J^1\pi)$ is the
{\sl Lagrangian function} determined by $\Lden$. Throughout this paper
we denote by $\d t$ the volume form in $\Real$, and its pull-backs
to all the manifolds.

The {\sl Poincar\'e-Cartan forms} associated with the
Lagrangian density $\Lden$ are defined using the
{\sl vertical endomorphism} ${\cal V}$ of the bundle
$J^1\pi$ (see \cite{EMR-91,Sa-89})
$$
\Theta_{\Lden}=\inn({\cal V})\d\Lden+\Lden\in\df^1(J^1\pi) \quad
;\quad \Omega_{\Lden}= -\d\Theta_{\Lden}\in\df^2(J^1\pi) \, .
$$
A Lagrangian $\Lden$ is {\sl regular} if $\Omega_{\Lden}$ has
maximal rank; elsewhere $\Lden$ is singular. In natural
coordinates we have \(\displaystyle {\cal V}=(\d q^i-v^i\d
t)\otimes\derpar{}{v^i} \otimes\derpar{}{t}\), and \beann
\Theta_{\Lden}&=&\derpar {L}{v^i}\d q^i-\left(\derpar {L}
{v^i}v^i- L \right)\d t
\\
\Omega_{\Lden}&=& -\frac{\partial^2 L }{\partial v^j\partial
v^i}\d v^j\wedge\d q^i- \frac{\partial^2 L }{\partial q^j\partial
v^i}\d q^j\wedge\d q^i
  \\  & &
+ \frac{\partial^2 L }{\partial v^j\partial v^i}v^i\d v^j\wedge\d
t  + \left(\frac{\partial^2 L }{\partial q^j\partial v^i}v^i
-\derpar {L} {q^j}+ \frac{\partial^2 L }{\partial t\partial
v^j}\right)\d q^j\wedge\d t\; .
 \eeann
  The regularity condition is equivalent to
\(\displaystyle det\left(\frac{\partial^2 L}{\partial v^i\partial
v^j}(\bar y)\right)\not= 0\), for every $\bar y\in J^1\pi$.
Geometrically,  $\Lden$ is regular if and only if
$(\Omega_{\Lden}, \d t)$ is a cosymplectic structure on $J^1\pi$.
This means that $\Omega_{\Lden}$ and $\d t$ are closed and
$\Omega_{\Lden}^{n} \wedge \d t$ is a volume form (see \cite{LR}).

The {\sl Lagrangian problem} consists in finding sections
$\phi\colon\Real\to E$ of $\pi$, characterized by
  $$
  (j^1\phi)^*\inn (X)\Omega_\Lden=0 \quad ,\quad
  \mbox{\rm for every $X\in\vf(J^1\pi)$}
  $$
where $j^1\phi: \R\to J^1\pi$ is the 1-jet
extension of $\phi$.
  In natural coordinates, if $\phi(t)=(t,\phi^i(t))$,
  this condition is equivalent to demanding that
  $\phi$ satisfies the {\sl Euler-Lagrange equations}
  $$
  \derpar {L} {q^i}\Big\vert_{j^1\phi}-
\frac{d}{d t}\left(\derpar {L} {v^i}\right) \Big\vert_{j^1\phi} = 0
  \quad , \quad \mbox{\rm (for $i=1,\ldots ,n$)}
  $$
where $j^1\phi(t)=(t,\phi^i(t), \dot{\phi}^i(t))$. Assuming that
these sections are integral curves of vector fields in $J^1\pi$
the corresponding equations for these vector fields are
 \beq \inn
(X_{\Lden})\Omega_{\Lden}=0 \quad , \quad \inn (X_{\Lden})\d t=1
\label{ELvf}
 \eeq
 where $X_{\Lden}\in\vf(J^1\pi)$ is holonomic
(recall that a vector field in $J^1\pi$ is said to be holonomic,
or also a {\sl second order differential equation} (SODE for
simplicity), if its integral curves are holonomic; that is,
canonical liftings of sections $\varphi\colon\Real\to E$). In the
regular case, there is a unique solution to these equations. In
the singular case the existence of a solution is not assured,
except perhaps on some submanifold (or subset)
of $J^1\pi$, where the solution is not unique, in general.

Consider now the {\sl extended momentum phase space} $\Tan^*E$,
and the {\sl restricted momentum phase space} which is defined by
$J^1\pi^*= \Tan^*E/\pi^*\Tan^*\Real$. Local coordinates in these
manifolds are $(t,q^i,p,p_i)$ and $(t,q^i,p_i)$, respectively.
Then, the following natural projections are
$$
%\beann
\tau^1 \colon J^1\pi^*\to E \quad ,\quad \bar\tau^1=\pi\circ\tau^1
\colon J^1\pi^* \to \Real \quad , \quad \mu \colon\Tan^*E \to
J^1\pi^* \quad , \quad p\colon\Tan^*E \to \Real \, .
%\eeann
$$
Let $\Theta\in\df^1(\Tan^*E)$ and
$\Omega=-\d\Theta\in\df^2(\Tan^*E)$ be
the canonical forms of $\Tan^*E$ whose local expressions are
$$
\Theta=p_i\d q^i+ p\d t \quad ,\quad \Omega=\d q^i\wedge\d p_i+\d
t\wedge\d p \, .
$$
(In the particular case $E=\Real\times Q$, we have $\Tan^*E \simeq
\Real \times \Real^*\times\Tan^*Q$, and $J^1\pi^* \simeq
\Real\times\Tan^*Q$ and introducing the projections $pr_1
\colon\Tan^*(\Real\times Q) \to\Real\times\Real^*$, $pr_2
\colon\Tan^*(\Real\times Q) \to \Tan^*Q$, we have
$\Theta = pr_1^*\Theta_{_{\Real}} + pr_2^*\Theta_Q$ and $\Omega =
pr_1^*\Omega_{_{\Real}}+ pr_2^*\Omega_Q$; where
$\Omega_{_{\Real}}=-\d\Theta_{_{\Real}} \in\df^2(\Real \times
\Real^*)$ and $\Omega_Q=-\d\Theta_Q \in \df^2(\Tan^*Q)$
 denote the natural symplectic forms of $\Real \times \Real^*$ and $\Tan^*Q$).

Being $\Theta_{\Lden}\in\df^1(J^1\pi)$ $\pi^1$-semibasic,
we have a natural map
$\widetilde{{\cal F}\Lden}\colon J^1\pi\to\Tan^*E$,
given by
   \beq
  \widetilde{{\cal F}\Lden}({\bar y})=\Theta_{\Lden}({\bar y})
  \label{elm}
  \eeq
which is called the {\sl extended Legendre map} associated to the
Lagrangian density $\Lden$. The {\sl restricted Legendre map} is
${\cal F}\Lden =\mu\circ\widetilde{{\cal F}\Lden}\colon J^1\pi\to
J^1\pi^*$. Their local expressions are
  \beann
  \widetilde{{\cal F}\Lden}^*t = t \quad\ , \ \quad
  \widetilde{{\cal F}\Lden}^*q^i = q^i \quad\  , \quad
  \widetilde{{\cal F}\Lden}^*p_i =\derpar {L} {v^i} &\  , \ &
  \widetilde{{\cal F}\Lden}^*p = L -v^i\derpar {L} {v^i}
  \\
  {\cal F}\Lden^*t = t \quad\ , \ \quad
  {\cal F}\Lden^*q^i = q^i \quad\  , \quad
  {\cal F}\Lden^*p_i =\derpar {L} {v^i} & &
  \eeann
  or, in other words, $\widetilde{{\cal F}\Lden}(t, q^i,
  \dot{q}^i)=(t, q^i, L -v^i\derpar {L} {v^i}, \derpar {L} {v^i})$
  and ${{\cal F}\Lden}(t, q^i,
  \dot{q}^i)=(t, q^i, \derpar {L} {v^i})$. Moreover, we have $\widetilde{{\cal F}\Lden}^*\Theta=\Theta_{\Lden}$,
and $\widetilde{{\cal F}\Lden}^*\Omega=\Omega_{\Lden}$.

%\subsection{The hyper-regular and regular cases}

 The Lagrangian $\Lden$ is regular
 if, and only if, ${\cal F}\Lden$ is a local diffeomorphism.
 As a particular case, $\Lden$ is a {\sl hyper-regular}
 Lagrangian if ${\cal F}\Lden$ is a global diffeomorphism.

If $\Lden$ is a hyper-regular Lagrangian, then
  $\tilde{\cal P}=\widetilde{{\cal F}\Lden}(J^1\pi)$ is a
  1-codimensional, $\mu$-transverse embedded submanifold of $\Tan^*E$,
  with natural embedding $\tilde\jmath_0\colon\tilde{\cal P}\hookrightarrow\Tan^*E$,
  which is diffeomorphic to $J^1\pi^*$.
This diffeomorphism is the inverse of $\mu$
 restricted to $\tilde{\cal P}$, and also coincides with the map
  $h=\widetilde{{\cal F}\Lden}\circ{\cal F}\Lden^{-1}$,
  when it is restricted onto its image (which is just $\tilde{\cal P}$).
  This map $h$ is called a {\sl Hamiltonian section}, and
  is used to construct the {\sl Hamilton-Cartan forms}
  in $J^1\pi^*$ by making
  $$
  \Theta_h=h^*\Theta\in\df^1(J^1\pi^*)
\quad , \quad \Omega_h=h^*\Omega\in\df^2(J^1\pi^*) \, .
  $$
Locally, the Hamiltonian section $h$ is specified by
$h(t,q^i,p_i)=(t,q^i,-H,p_i)$, where $H$ is
the local Hamiltonian function given by
  $H=p_i(F\Lden^{-1})^*v^i-(F\Lden^{-1})^* L $.
  The local expressions are
  $$
  \Theta_h = p_i\d q^i-H\d t  \quad , \quad
  \Omega_h = \d q^i\wedge\d p_i +\d H\wedge\d t \, .
  $$
   Of course
  ${\cal F}\Lden^*\Theta_h=\Theta_{\Lden}$,
  and ${\cal F}\Lden^*\Omega_h=\Omega_{\Lden}$.

  The {\sl Hamiltonian problem} consists in finding
  sections of $\bar\tau^1$, $\psi\colon\Real\to J^1\pi^*$,
  characterized~by
  $$
  \psi^*\inn (X)\Omega_h=0 \quad , \quad
  \mbox{\rm  for every $X\in\vf (J^1\pi^*)$\ .}
  $$
 This condition leads to the {\sl Hamilton equations} which, if
  $\psi(t)=(t,q^i(t),p_i(t))$, in natural coordinates are
$$
\frac{d q^i}{d t}=\derpar{H}{p_i}\Bigg\vert_{\psi} \quad ; \quad
\frac{d p_i}{d t}=-\derpar{H}{q^i}\Bigg\vert_{\psi} \, .
  $$
Assuming that these sections are integral curves of vector fields $X_h\in\vf(J^1\pi^*)$,
the corresponding equations for these vector fields are
$$
\inn (X_h)\Omega_h=0 \quad , \quad \inn (X_h)\d t=1\, .
$$

As a final remark, it can be proved that solutions to the Lagrangian
and Hamiltonian problems are equivalent, in the sense that they are
${\cal F}\Lden$-related; that is,
 \beq
 \psi={\cal F}\Lden\circ
j^1\phi \quad ; \quad \Tan {\cal F}\Lden\circ X_{\Lden}=X_h\circ
{\cal F}\Lden \, .
\label{equivsol}
 \eeq

For regular, but not hyper-regular systems, the results are
the same, but only locally on open neighbourhoods at every point,
instead of $J^1\pi^*$.

%\subsection{The almost-regular case}

A singular Lagrangian $\Lden$ is {\sl almost-regular} if:
  ${\cal P}={\cal F}\Lden ( J^1\pi)$
  is a closed submanifold of $J^1\pi^*$
  (let $\jmath\colon {\cal P}\hookrightarrow J^1\pi^*$ be natural embedding),
  ${\cal F}\Lden$ is a submersion onto its image, and
  for every $\bar y\in  J^1\pi$, the fibres
  ${\cal F}\Lden^{-1}({\cal F}\Lden (\bar y))$
  are connected submanifolds of $J^1\pi$.

  If $\Lden$ is an almost-regular Lagrangian,
the submanifold ${\cal P}$ of $J^1\pi^*$
is a fibre bundle over $E$ and $M$. In this case the
$\mu$-transverse submanifold $\tilde\jmath\colon \tilde{\cal
P}\hookrightarrow\Tan^*E$ is diffeomorphic to ${\cal P}$. This
diffeomorphism is denoted by $\tilde\mu\colon\tilde{\cal
P}\to{\cal P}$, and is just the restriction of the projection
$\mu$ to $\tilde{\cal P}$. Then, taking the Hamiltonian section
$\tilde h=\tilde\jmath\circ\tilde\mu^{-1}$, we define the forms
  $$
\Theta^0_h=\tilde h^*\Theta
\quad ; \quad
\Omega^0_h=\tilde h^*\Omega
$$
which verify that
${\cal F}\Lden_0^*\Theta^0_h=\Theta_{\Lden}$ and
${\cal F}\Lden_0^*\Omega^0_h=\Omega_{\Lden}$
(where ${\cal F}\Lden_0$ is the restriction map of
${\cal F}\Lden$ onto ${\cal P}$).
%Then we have the following diagram
%$$\bfig\xymatrix{&& \tilde{\cal P}\ar[rr]_{\txt{\small{$\tilde\jmath$}}}
%\ar[d]^{\txt{\small{$\tilde\mu$}}} && T^*E
%\ar[d]_{\txt{\small{$\mu$}}}
% \\ J^1\pi \ar[urr]^{\txt{\small{$\widetilde{{\cal F}\Lden}_0$}}}
% \ar[rr]_{\txt{\small{${\cal F}\Lden_0$}}}
%&& {\cal P} \ar@<4pt>[u]^{\txt{\small{$\tilde\mu^{-1}$}}}
%\ar[urr]_{\txt{\small{$\tilde h$}}}
%\ar[rr]_{\txt{\small{$\jmath$}}}\ar[dr]_{\txt{\small{$\bar\tau^1_0$}}}&&
%J^1\pi^*
%\ar[dl]^{\txt{\small{$\bar\tau^1$}}} \\
%&& & \Real & }\efig \label{arhs}
%$$
  Then, the Hamiltonian problem and the equations of motion
  are stated as in the hyper-regular case.
  Now, the existence of a solution to these equations is not assured,
  except perhaps on some submanifold
  of ${\cal P}$, where the solution is not unique, in general.

\section{Unified formalism}
\protect\label{uf}

We define the {\sl extended jet-momentum bundle} ${\cal W}$ and
the {\sl restricted jet-momentum bundle} ${\cal W}_r$
$$
{\cal W}= J^1\pi \times_{E}\Tan^*E
\quad , \quad
{\cal W}_r =  J^1\pi \times_{E}  J^1\pi^*
$$
with natural coordinates $(t,q^i,v^i,p,p_i)$ and $(t,q^i,v^i,p_i)$,
respectively. Natural submersions are
 \bea \rho_1\colon{\cal
W}\to  J^1\pi \ ,\ \rho_2\colon{\cal W}\to \Tan^*E \ ,\
\rho_{_{E}}\colon{\cal W}\to E \ ,\ \rho_{_{\Real}}\colon{\cal W}\to
\Real \label{project}
\\
\rho_1^r\colon{\cal W}_r\to  J^1\pi \ ,\ \rho_2^r\colon{\cal W}_r\to
J^1\pi^* \ ,\ \rho_{_{E}}^r\colon{\cal W}_r\to E \ ,\
\rho_{_{\Real}}^r\colon{\cal W}_r\to \Real \, ,\nonumber
\eea
  with $\pi^1\circ\rho_1=\tau^1\circ\mu\circ\rho_2=\rho_{_{E}}$.
For $\bar y\in J^1\pi$, ${\bf p}\in\Tan^*E$, and
$[{\bf p}]=\mu({\bf p})\in J^1\pi^*$,
there is also the natural projection
$$
\begin{array}{ccccc}
\mu_{_{\cal W}}&\colon& {\cal W}& \to &{\cal W}_r\\
& & (\bar y,{\bf p}) & \mapsto & (\bar y,[{\bf p}])
\end{array}
$$

The bundle ${\cal W}$ is endowed with the following canonical structures:

\begin{definition}
\ben
\item
The {\rm coupling $1$-form} in ${\cal W}$ is the
$\rho_{_{\Real}}$-semibasic $1$-form $\hat{\cal C}\in\df^1({\cal
W})$ defined as follows: for every
$w=(j^1\phi(t),\alpha)\in {\cal W}$ (that is, $\alpha\in
T^*_{\rho_{_E}(w)} E$) and  $V\in \Tan_w{\cal W}$, then
$$
\hat{\cal C}(V)=\alpha(\Tan_w(\phi\circ\rho_{_{\Real}})V)\; .
$$
\item
The {\rm canonical $1$-form}
$\Theta_{\cal W}\in\df^1({\cal W})$ is the $\rho_{_{E}}$-semibasic form defined by
$\Theta_{\cal W}=\rho_2^*\Theta$.

The {\rm canonical $2$-form} is
$\Omega_{\cal W}=-\d\Theta_{\cal W}=\rho_2^*\Omega\in\df^2({\cal W})$.
\een
\label{coupling}
\end{definition}

Being $\hat{\cal C}$ a $\rho_{_{\Real}}$-semibasic form,
there is $\hat C\in\Cinfty ({\cal W})$ such that $\hat{\cal C}=\hat C\d t$.
 Note also that $\Omega_{\cal W}$ is degenerate, its kernel being the $\rho_2$-vertical vectors;
then $({\cal W},\Omega_{\cal W} )$ is a presymplectic manifold.

The local expressions for $\Theta_{\cal W}$, $\Omega_{\cal W}$, and $\hat{\cal C}$ are
$$
\Theta_{\cal W} = p_i\d q^i+p\d t  \quad , \quad
  \Omega_{\cal W} = -\d p_i\wedge\d q^i-\d p\wedge\d t   \quad , \quad
\hat{\cal C}=(p+p_i v^i)\d t \, .
$$

Given a Lagrangian density $\Lden\in\df^1( J^1\pi)$, we denote
$\hat\Lden=\rho_1^*\Lden\in\df^1({\cal W})$, and we can write
$\hat\Lden=\hat  L\d t$, with $\hat
 L =\rho_1^* L \in\Cinfty ({\cal W})$. We define a {\sl Hamiltonian
submanifold}
$$
{\cal W}_0=\{ w\in{\cal W}\ | \ \hat\Lden(w)=\hat{\cal C}(w) \}\, .
$$
So, ${\cal W}_0$ is the submanifold of $\cal W$ defined by the
regular constraint function $\hat C-\hat  L =0$, which
is globally defined in $\cal W$
using the dynamical data and the geometry.
In local coordinates it is
 $$
\hat C-\hat  L =p + p_i v^i-\hat  L (t,q^j,v^j)=0 \ .
 $$
The natural embedding is
$\jmath_0\colon{\cal W}_0\hookrightarrow{\cal W}$.
We have the projections (submersions), see diagram~(\ref{diag0}):
$$
\rho_1^0\colon{\cal W}_0\to  J^1\pi \ ,\
\rho_2^0\colon{\cal W}_0\to \Tan^*E \ ,\
\rho_{_{E}}^0\colon{\cal W}_0\to E \ ,\
\rho_{_{\Real}}^0\colon{\cal W}_0\to \Real
$$
which are the restrictions to ${\cal W}_0$ of the projections
(\ref{project}), and
 $$
\hat {\rho }_2^0 = \mu\circ \rho _2^0\colon{\cal W}_0 \to J^1\pi^* \, .
$$
Local coordinates in ${\cal W}_0$ are
$(t,q^i,v^i,p_i)$, and we have that
$$
\begin{array}{ccc}
\rho_1^0(t,q^i,v^i,p_i) = (t,q^i,v^i)  & , &
\jmath_0(t,q^i,v^i,p_i) = (t,q^i,v^i, L -v^i p_i,p_i) \\
\hat\rho_2^0(t,q^i,v^i,p_i) = (t,q^i,p_i) & , &
\rho_2^0(t,q^i,v^i,p_i) =(t,q^i, L -v^i p_i,p_i) \, .
\end{array}
$$

\begin{prop}
${\cal W}_0$ is a $1$-codimensional $\mu_{_{\cal W}}$-transverse
submanifold
of ${\cal W}$, diffeomorphic to~${\cal W}_r$.
\label{1}
\end{prop}
\proof
For every $(\bar y,{\bf p})\in{\cal W}_0$, we have
$ L (\bar y)\equiv\hat  L (\bar y,{\bf p})=\hat C(\bar y,{\bf p})$, and
$$
(\mu_{_{\cal W}}\circ\jmath_0)(\bar y,{\bf p})= \mu_{_{\cal W}}(\bar y,{\bf p})=(\bar y,\mu({\bf p})) \ .
$$

First, $\mu_{_{\cal W}}\circ\jmath_0$ is injective:
let $(\bar y_1,{\bf p}_1),(\bar y_2,{\bf p}_2)\in{\cal W}_0$, then we have
$$
(\mu_{_{\cal W}}\circ\jmath_0)(\bar y_1,{\bf p}_1)=
(\mu_{_{\cal W}}\circ\jmath_0)(\bar y_2,{\bf p}_2)
\,\Rightarrow\, (\bar y_1,\mu({\bf p}_1))=(\bar y_2,\mu({\bf p}_2))
\, \Rightarrow\,
\bar y_1=\bar y_2\ ,\ \mu({\bf p}_1)=\mu({\bf p}_2)
$$
hence
$ L (\bar y_1)= L (\bar y_2)=\hat C(\bar y_1,{\bf p}_1)=
\hat C(\bar y_2,{\bf p}_2)$.
In a local chart, the third equality gives
$$
p({\bf p}_1)+p_i({\bf p}_1)v^i(\bar y_1)=
p({\bf p}_2)+p_i({\bf p}_2)v^i(\bar y_2)
$$
but $\mu({\bf p}_1)=\mu({\bf p}_2)$ implies
$p_i({\bf p}_1)=p_i([{\bf p}_1])=p_i([{\bf p}_2])=p_i({\bf p}_2)$;
then $p({\bf p}_1)=p({\bf p}_2)$, and~${\bf p}_1=~{\bf p}_2$.

Second, $\mu_{_{\cal W}}\circ\jmath_0$ is onto, then, if $(\bar y,[{\bf
p}])\in{\cal W}_r$, there exists $(\bar y,{\bf
q})\in\jmath_0({\cal W}_0)$ such that $[{\bf q}]=[{\bf p}]$. In
fact, it suffices to take $[{\bf q}]$ such that, in a
local chart of $ J^1\pi \times_E\Tan^*E={\cal W}$
$$
p_i({\bf q})=p_i([{\bf p}]) \ , \ p({\bf q})= L (\bar y)-p_i([{\bf
p}])v^i(\bar y) \, .
$$

Finally, since ${\cal W}_0$ is defined by the constraint function
$\hat C-\hat  L $ and, as \(\ker\,\mu_{\cal W*}=\left\{\derpar{}{p}\right\}\)
 locally and
\(\displaystyle\derpar{}{p}(\hat C-\hat  L )=1\), then ${\cal W}_0$
is $\mu_{_{\cal W}}$-transversal.
 \qed

As a consequence of this result, the
submanifold ${\cal W}_0$ induces a section
$\hat h\colon{\cal W}_r\to{\cal W}$ of the projection $\mu_{_{\cal W}}$.
Locally, $\hat h$ is specified by giving the local {\sl
Hamiltonian function} $\hat H= -\hat  L +p_i v^i$; that is, $\hat
h(t,q^i,v^i,p_i)=(t,q^i,v^i,-\hat H,p_i)$. In this sense, $\hat h$
is a {\sl Hamiltonian section} of $\mu_{_{\cal W}}$.

So we have the following diagram
\beq\bfig\xymatrix{&& J^1\pi &&\\
{\cal W}_0 \ar[urr]^{\txt{\small{$\rho_1^0$}}}
\ar[rr]^{\txt{\small{$\jmath_0$}}}
\ar[drr]^{\txt{\small{$\rho_2^0$}}}
\ar[ddrr]_{\txt{\small{$\hat\rho_2^0$}}}&& {\cal W}
\ar[u]^{\txt{\small{$\rho_1$}}}
\ar[d]_{\txt{\small{$\rho_2$}}}\ar[rr]^{\txt{\small{$\mu_{_{\cal
W}}$}}} && {\cal W}_r \ar[ull]_{\txt{\small{$\rho_1^r$}}}
\ar[dll]_{\txt{\small{$\rho_2\circ\hat h$}}}
\ar[ddll]^{\txt{\small{$\rho_2^r$}}}\\ && T^*E
\ar[d]_{\txt{\small{$\mu$}}}&& \\ && J^1\pi^* && }\efig
\label{diag0} \eeq

\textbf{Remark}: Observe that, from the Hamiltonian $\mu_{_{\cal
W}}$-section $\hat h\colon{\cal W}_r\to{\cal W}$ in the extended
unified formalism, we can recover the Hamiltonian $\mu$-section
$\tilde h=\tilde\jmath\circ\tilde\mu^{-1}\colon {\cal
P}\to\Tan^*E$ in the standard Hamiltonian formalism assuming that
${\mathcal L}$ is almost-regular. In fact, given $[{\bf p}]\in
J^1\pi^*$, the section $\hat h$ maps every point $(\bar y,[{\bf
p}])\in(\rho_2^r)^{-1}([{\bf p}])$ into $\rho_2^{-1}[\rho_2(\hat
h(\bar y,[{\bf p}]))]$. Now, the crucial point is the
projectability of the local function $\hat H$ by $\rho_2$. However,
\(\displaystyle\derpar{}{v^i}\) being a local basis for
$\ker\,\rho_{2*}$, $\hat H$ is $\rho_2$-projectable if, and only
if,
  \(\displaystyle p_i=\derpar {L} {v^i}\), and this condition
  is fulfilled when
  $[{\bf p}]\in{\cal P}={\rm Im}\,{\cal F}\Lden\subset J^1\pi^*$,
  which implies that
  $\rho_2[\hat h((\rho_2^r)^{-1}([{\bf p}]))]\in
  \tilde{\cal P}={\rm Im}\, \widetilde{{\cal F}\Lden}\subset\Tan^*E$.
  Then, the Hamiltonian section $\tilde h$ is defined as
  $$
\tilde h([{\bf p}])=(\rho_2\circ\hat
h)[(\rho_2^r)^{-1}(\jmath([{\bf p}]))]=
(\tilde\jmath\circ\tilde\mu^{-1})([{\bf p}]) \ ,\ \mbox{\rm for
every $[{\bf p}]\in{\cal P}$}.
$$
So we have the diagram
$$
\bfig\xymatrix{\tilde{\cal P}\ar[rr]_{\txt{\small{$\tilde\jmath$}}}
\ar[d]^{\txt{\small{$\tilde\mu$}}}&& T^*E \ar[d]^{\txt{\small{$\mu$}}}
&& {\cal W} \ar[ll]^{\txt{\small{$\rho_2$}}} \\
{\cal P} \ar[u]<4pt>^{\txt{\small{$\tilde\mu^{-1}$}}}
\ar[urr]_{\txt{\small{$\tilde h$}}}
\ar[rr]_{\txt{\small{$\jmath$}}}&& J^1\pi^* && {\cal W}_r
\ar[ll]^{\txt{\small{$\rho^r_2$}}}\ar[u]_{\txt{\small{$\hat h$}}}
}\efig
$$
For (hyper) regular systems this diagram is the same with ${\cal
P}={\rm Im}\,{\cal F}\Lden=J^1\pi^*$.

Finally, we can define the forms
  $$
\Theta_0=\jmath_0^*\Theta_{\cal W} = \rho _2^{0*} \Theta \in\df^1({\cal W}_0)
\quad , \quad
\Omega_0=\jmath_0^*\Omega_{\cal W} =\rho _2^{0*} \Omega
\in\df^2({\cal W}_0)
$$
with local expressions
\beq
\Theta_0=( L -p_i v^i)\d t+p_i\d q^i
\quad, \quad
\Omega_0=\d (p_i v^i- L )\wedge\d t-\d p_i\wedge\d q^i
\label{coor0}
\eeq
and we have the presymplectic Hamiltonian systems
$({\cal W}_0,\Omega_0)$ and $({\cal W}_r,\Omega_r)$,
with $\Omega_r=\hat h^*\Omega_0$.

\section{The dynamical equations for sections}
\protect\label{des}

Now we establish the dynamical problem for
the system $({\cal W}_0,\Omega_0)$ which,
as a consequence of the diffeomorphism stated
in Proposition \ref{1}, is equivalent to making it
for the system $({\cal W}_r,\Omega_r)$.

The {\sl Lagrange-Hamiltonian problem} associated with the system
$({\cal W}_0,\Omega_0)$ consists in finding sections of
$\rho^0_{_{\Real}}$, $\psi_0\colon\Real\to{\cal W}_0$, which are
characterized by the condition
  \beq
  \psi_0^*\inn(Y_0)\Omega_0=0 \quad ,\quad
  \mbox{\rm for every $Y_0\in\vf({\cal W}_0)$} \ .
  \label{psi0}
\eeq
This equation gives different kinds of information, depending
on the type of the vector fields $Y_0$ involved. In particular,
using  $\hat\rho_2^0$-vertical vector fields,
denoted by $\vf^{{\rm V}(\hat\rho_2^0)}({\cal W}_0)$, we have:

\begin{lem}
If $Y_0\in\vf^{{\rm V}(\hat\rho_2^0)}({\cal W}_0)$,
then $\inn(Y_0)\Omega_0$ is $\rho_\Real^0$-semibasic.
\end{lem}
\proof
A simple calculation in coordinates leads
to this result. In fact,
taking \(\left\{\derpar{}{v^i}\right\}\)
as a local basis for the $\hat\rho_2^0$-vertical vector fields,
and bearing in mind (\ref{coor0}) we obtain
the $\rho_{_{\Real}}^0$-semibasic forms
$$
\inn\left(\derpar{}{v^i}\right)\Omega_0=
\left( p_i-\derpar {L} {v^i}\right)\d t \ .
$$
\qed

As an immediate consequence, when
 $Y_0\in\vf^{{\rm V}(\hat\rho_2^0)}({\cal W}_0)$, condition (\ref{psi0}) does not
depend on the derivatives of $\psi_0$: it is a pointwise
(algebraic) condition. We can define the submanifold
$$
{\cal W}_1 = \{ (\bar y,{\bf p})\in{\cal W}_0 \ | \
\inn(V_0)(\Omega_0)_{({\bar y},{\bf p})}=0,\
\mbox{for every $V_0\in{\rm V}_{({\bar y},{\bf p})}(\hat\rho_2^0 )$} \}
$$
where ${\rm V}(\hat\rho_2^0 )$ denotes the $\hat\rho_2^0$-vertical
vectors. ${\cal W}_1$ is called the {\sl first constraint
submanifold} of the Hamiltonian pre-multisymplectic system $({\cal
W}_0,\Omega _0)$, as every section $\psi_0$ solution to
(\ref{psi0}) must take values in  ${\cal W}_1$. We denote by
$\jmath_1\colon{\cal W}_1\hookrightarrow{\cal W}_0$ the natural
embedding.

Locally, ${\cal W}_1$ is defined in ${\cal W}_0$ by the constraints
\(\displaystyle p_i=\derpar {L} {v^i}\). Moreover:

\begin{prop}
${\cal W}_1$ is the graph of $\widetilde{{\cal F}\Lden}$; that is,
${\cal W}_1=\{ (\bar {y},\widetilde{{\cal F}\Lden}(\bar y))\in{\cal W}\  \mid \
\bar y\in  J^1\pi\}$.
\label{transve}
\end{prop}
\proof
 Consider $\bar y\in  J^1\pi$, let $\phi\colon\Real\to E$
be a representative of $\bar y$, and ${\bf p}=\widetilde{{\cal F}\Lden}(\bar y)$.
For every $U\in\Tan_{\bar\pi^1(\bar y)}\Real$,
consider $V=\Tan_{\bar\pi^1(\bar y)}\phi(U)$ and its canonical
lifting $\bar V=\Tan_{\bar\pi^1(\bar y)}j^1\phi(U)$. From the
definition of the extended Legendre map (\ref{elm}) we have
$(\Tan_{\bar y}\pi^1)^*(\widetilde{{\cal F}\Lden}(\bar y))=(\Theta_\Lden)_{\bar y}$, then
$$
\inn(\bar V)[(\Tan_{\bar y}\pi^1)^*(\widetilde{{\cal F}\Lden}(\bar
y))]= \inn(\bar V)(\Theta_\Lden)_{\bar y} \, .
$$
Furthermore, as ${\bf p}=\widetilde{{\cal F}\Lden}(\bar y)$, we also
have that \beann \inn(\bar V)[(\Tan_{\bar
y}\pi^1)^*(\widetilde{{\cal F}\Lden}(\bar y))]&=&
\inn(\Tan_{\bar\pi^1(\bar y)}j^1\phi(U))[(\Tan_{\bar y}\pi^1)^*{\bf
p}]= \inn((\Tan_{\bar y}\pi^1)_*(\Tan_{\bar\pi^1({\bar
y})}j^1\phi(U))){\bf p} \nonumber \\ &=& \inn(\Tan_{\bar\pi^1(\bar
y)}\phi(U)){\bf p}=\inn(V){\bf p} \, .
\eeann
Therefore we obtain
$$
\inn(U)(\phi^*{\bf p})=\inn(U)[(j^1\phi)^*(\Theta_\Lden)_{\bar y}]
$$
and bearing in mind the definition of the coupling form $\cal C$,
this condition becomes
$$
\inn(U)(\hat{\cal C}(\bar y,{\bf
p}))=\inn(U)[(j^1\phi)^*\Theta_\Lden)_{\bar y}] \, .
$$
Since it holds for every $U\in\Tan_{\bar\pi^1(\bar y)}\Real$,
we conclude that
$\hat{\cal C}(\bar y,{\bf p})=[(j^1\phi)^*\Theta_\Lden]_{\bar y}$,
or equivalently, $\hat{\cal C}(\bar y,{\bf p})=\hat  L ({\bar y,{\bf p})}$,
where we have made use of the fact that $\Theta_\Lden$ is the sum
of the Lagrangian density $\Lden$ and a contact form
$\inn({\cal V})\d\Lden$ (vanishing by pull-back of lifted sections).
This is the condition defining ${\cal W}_0$,
and thus we have proved that $(\bar y,\widetilde{{\cal F}\Lden}(\bar y))\in{\cal W}_0$,
for every $\bar y\in  J^1\pi$; that is,
${\rm graph}\,\widetilde{{\cal F}\Lden}\subset{\cal W}_0$.
Furthermore, ${\rm graph}\,\widetilde{{\cal F}\Lden}$ and ${\cal W}_1$
are defined as subsets of ${\cal W}_0$ by the same local conditions:
\(\displaystyle p_i -\derpar {L} {v^i} =0\).
So we conclude that
${\rm graph}\,\widetilde{{\cal F}\Lden}={\cal W}_1$.
\qed

As ${\cal W}_1$ is the graph of $\widetilde{{\cal F}\Lden}$, it is diffeomorphic
to $ J^1\pi$. Every section $\psi_0\colon\Real \to {\cal W}_0$ is of the form
$\psi_0=(\psi_\Lden,\psi_{\cal H})$, with
$\psi_\Lden=\rho_1^0\circ\psi_0\colon\Real \to  J^1\pi$,
and if $\psi_0$ takes values in ${\cal W}_1$ then
$\psi_{\cal H}=\widetilde{{\cal F}\Lden}\circ\psi_\Lden\colon\Real \to \Tan^*E$.
In this way every constraint, differential equation, etc.
in the unified formalism can be translated to the
Lagrangian or the Hamiltonian formalisms by restriction to the first
or the second factors of the product bundle.

However, as was pointed out before, the geometric condition
(\ref{psi0}) in ${\cal W}_0$, which can be solved only for
sections $\psi_0\colon\Real\to{\cal W}_1\subset{\cal W}_0$, is
stronger than the Lagrangian condition
$\psi_\Lden^*\inn(Z)\Omega_\Lden= 0$, (for every $Z\in\vf( J^1\pi)$) in
$ J^1\pi$, which can be translated to ${\cal W}_1$ by the natural
diffeomorphism between them. The reason is that,
as $\rho_1^0$ is a submersion,
and ${\cal W}_1$ is a $\rho_1^0$-transversal submanifold of ${\cal W}_0$
(as a consequence of Proposition \ref{transve}),
we have the splitting
$\jmath_1^*\Tan{\cal W}_0=\Tan {\cal W}_1\oplus_{{\cal W}_1}\jmath_1^*{\rm V}(\rho_1^0)$,
$\jmath_1\colon{\cal W}_1\hookrightarrow{\cal W}_0$ being the natural
embedding. Therefore the additional information comes
from the $\rho_1^0$-vertical vectors, and is just the holonomic
condition. In fact:

\begin{teor}
Let $\psi_0\colon\Real\to{\cal W}_0$ be a section fulfilling equation
(\ref{psi0}),
$\psi_0=(\psi_\Lden,\psi_{\cal H})=(\psi_\Lden,\widetilde{{\cal F}\Lden}\circ\psi_\Lden)$,
where $\psi_\Lden=\rho_1^0\circ\psi_0$. Then:
\ben
  \item
$\psi_\Lden$ is the canonical lift of the projected section
$\phi=\rho_E^0\circ\psi_0\colon\Real\to E$ (that is, $\psi_\Lden$ is
a holonomic section).
\item
The section $\psi_\Lden=j^1\phi$ is a
solution to the Lagrangian problem, and the section
$\mu\circ\psi_{\cal H}=
\mu\circ\widetilde{{\cal F}\Lden}\circ\psi_\Lden={\cal F}\Lden\circ j^1\phi$
is a solution to the Hamiltonian problem.
\een
Conversely, for every section $\phi\colon\Real\to E$ such that
$j^1\phi$ is a solution to the Lagrangian problem (and hence ${\cal F}\Lden\circ j^1\phi$
 is a solution to the Hamiltonian problem) we have
that the section $\psi_0=(j^1\phi,\widetilde{{\cal F}\Lden}\circ
j^1\phi)$, is a solution to (\ref{psi0}).
 \label{mainteor1}
\end{teor}
\proof
\ 1. \
%\ben
%\item
Taking \(\left\{\derpar{}{p_i}\right\}\)
as a local basis for the $\rho^0_1$-vertical vector fields:
$$
\inn\left(\derpar{}{p_i}\right)\Omega_0  = v^i\d t-\d q^i
$$
so that for a section $\psi_0$ we have
$$
0=\psi _0^*\left[\inn\left(\derpar{}{p_i}\right)\Omega_0\right]=
\left(v^i-\derpar{q^i}{t}\right)\d t
$$
and thus the holonomy condition appears naturally within the
unified formalism.
So we have that
\(\displaystyle\psi_0=\left(t,q^i,\frac{d q^i}{dt},\derpar {L} {v^i}\right)\),
since $\psi_0$ takes values in ${\cal W}_1$, and hence it is of
the form $\psi_0=(j^1\phi,\widetilde{{\cal F}\Lden}\circ j^1\phi)$, for
$\phi=(t,q^i)=\rho_{_{E}}^0\circ\psi_0$.
%\item

2. \
Consider the diagram
$$
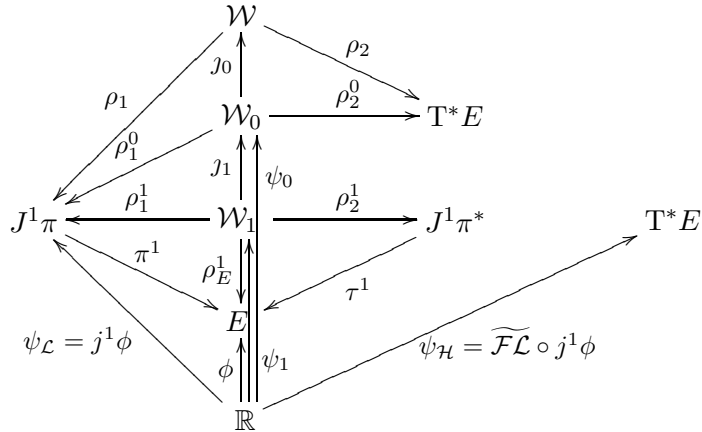
\bfig\xymatrix{&& {\cal W} \ar[ddll]_{\txt{\small{$\rho_1$}}}
\ar[drr]^{\txt{\small{$\rho_2$}}}&& &&\\ && {\cal W}_0 \ar[u]^{\txt{\small{$\jmath_0$}}}
\ar[dll]_{\txt{\small{$\rho^0_1$}}} \ar[rr]^{\txt{\small{$\rho^0_2$}}} && \Tan^*E && \\
J^1\pi\ar[drr]^{\txt{\small{$\pi^1$}}} && {\cal W}_1 \;
\ar[u]^{\txt{\small{$\jmath_1$}}}
\ar[ll]_{\txt{\small{$\rho^1_1$}}}
\ar[rr]^{\txt{\small{$\rho^1_2$}}}\ar[d]_{\txt{\small{$\rho^1_E$}}}&&
J^1\pi^*
\ar[dll]^{\txt{\small{$\tau^1$}}} && \Tan^*E \\
&& E \;  && && \\ && \; \Real
\ar[uull]^{\txt{\small{$\psi_\Lden=j^1\phi$}}}
\ar[u]^{\txt{\small{$\phi$}}}\ar[uu]<-3pt>_(.3){\txt{\small{$ \,
\psi_1$}}} \ar[uuu]<-6pt>_(.8){\txt{\small{$\psi_0$}}}
\ar[uurrrr]_{\txt{\small{$\; \psi_{\cal H}=\widetilde{{\cal F}\Lden}\circ j^1\phi$}}} && && }\efig
$$
Since sections
$\psi_0\colon\Real\to{\cal W}_0$ solution to (\ref{psi0}) take values
in ${\cal W}_1$, we can identify them with sections
$\psi_1\colon\Real\to{\cal W}_1$. These sections $\psi_1$ verify, in
particular,
that $\psi_1^*\inn(Y_1)\Omega_1=0$ holds for every
$Y_1\in\vf({\cal W}_1)$. Obviously $\psi_0=\jmath_1\circ\psi_1$.
Moreover, as ${\cal W}_1$ is the graph of $\widetilde{{\cal F}\Lden}$,
denoting by $\rho_1^1=\rho_1^0\circ\jmath_1\colon{\cal W}_1 \to
J^1\pi$ the diffeomorphism which identifies ${\cal W}_1$ with
$ J^1\pi$, if we define $\Omega_1=\jmath_1^*\Omega_0$, we have that
$\Omega_1=\rho_1^{1*}\Omega_\Lden$. In fact; as
$(\rho_1^1)^{-1}(\bar y)=(\bar y,\widetilde{{\cal F}\Lden}(\bar
y))$, for every $\bar y\in  J^1\pi$, then
$(\rho^0_2\circ\jmath_1\circ(\rho_1^1)^{-1})(\bar y)=
\widetilde{{\cal F}\Lden}(\bar y)\in\Tan^*E$, and hence
$$
\Omega_\Lden= (\rho^0_2\circ\jmath_1\circ(\rho_1^1)^{-1})^*\Omega=
[((\rho_1^1)^{-1})^*\circ\jmath_1^*\circ\rho^{0*}_2]\Omega=
[((\rho_1^1)^{-1})^*\circ\jmath_1^*]\Omega_0=((\rho_1^1)^{-1})^*\Omega_1
\, .
$$
Now, let $X\in\vf( J^1\pi)$. We have
\bea
(j^1\phi)^*\inn(X)\Omega_\Lden&=&
(\rho_1^0\circ\psi_0)^*\inn(X)\Omega_\Lden=
(\rho_1^0\circ\jmath_1\circ\psi_1)^*\inn(X)\Omega_\Lden
\nonumber \\ &=&
(\rho_1^1\circ\psi_1)^*\inn(X)\Omega_\Lden=
\psi_1^*\inn((\rho_1^1)_*^{-1}X)(\rho_1^{1*}\Omega_\Lden)=
\psi_1^*\inn(Y_1)\Omega_1
\nonumber \\ &=&
\psi_1^*\inn(Y_1)(\jmath_1^*\Omega_0)=
(\psi_1^*\circ\jmath_1^*)\inn(Y_0)\Omega_0=
\psi_0^*\inn(Y_0)\Omega_0
\label{chain}
\eea
where $Y_0\in\vf({\cal W}_0)$ is such that $Y_0=\jmath_{1*}Y_1$.
But as $\psi_0^*\inn(Y_0)\Omega_0=0$, for every $Y_0\in\vf({\cal W}_0)$,
then we conclude that $(j^1\phi)^*\inn(X)\Omega_\Lden=0$,
for every $X\in\vf( J^1\pi)$.
%\een

Conversely, let $j^1\phi\colon \Real\to  J^1\pi$ such that
$(j^1\phi)^*\inn(X)\Omega_\Lden=0$, for every $X\in\vf( J^1\pi)$,
and define $\psi_0\colon \Real\to{\cal W}_0$ as
$\psi_0=(j^1\phi,\widetilde{{\cal F}\Lden}\circ j^1\phi)$ (observe
that $\psi_0$ takes its values in ${\cal W}_1$). Taking into
account that, on the points of ${\cal W}_1$, every
$Y_0\in\vf({\cal W}_0)$ splits into $Y_0=Y_0^1+Y_0^2$, with
$Y_0^1\in\vf({\cal W}_0)$ tangent to ${\cal W}_1$, and
$Y_0^2\in\vf^{{\rm V}(\rho_1^0)}({\cal W}_0)$, we have that
$$
\psi_0^*\inn(Y_0)\Omega_0=\psi_0^*\inn(Y_0^1)\Omega_0+\psi_0^*\inn(Y_0^2)\Omega_0=0
$$
since for $Y_0^1$ the same reasoning as in (\ref{chain}) leads to
$$
\psi_0^*\inn(Y_0^1)\Omega_0=(j^1\phi)^*\inn(X_0^1)\Omega_\Lden=0
$$
(where $X_0^1=(\rho_1^1)_*Y_0^1$),
and for $Y_0^2$, following the same reasoning as in (\ref{chain}),
a local calculus gives
$$
\psi_0^*\inn(Y_0^2)\Omega_0=
(j^1\phi)^*\left[\left(f_i(x)\left(v_\alpha^A-\derpar{q^i}{x^\alpha}\right)\right)\d t\right]=0
$$
since $j^1\phi$ is a holonomic section and $\ds Y_0^2=f_i\derpar{}{p_i}$.
The result for the sections $\psi_{\cal H}=\widetilde{{\cal F}\Lden}\circ j^1\phi$
is a direct consequence of the first equivalence relations
(\ref{equivsol}).
\qed

{\bf Remark}:
The results in this section can also be recovered in coordinates
taking an arbitrary local vector field
\(\displaystyle Y_0=f\derpar{}{t}+f^i\derpar{}{q^i}+g^i\derpar{}{v^i}+
h_i\derpar{}{p_i}\in\vf({\cal W}_0)\), then
\beann
\inn(Y_0)\Omega_0 &=&-f\left(p_i\d v^i+v^i\d p_i-\derpar{L}{q^i}\d q^i-\derpar{L}{v^i}\d v^i\right)
\\ & &
-f^i\left(\derpar {L} {q^i}\d t+\d p_i\right) +
g^i\left(p_i-\derpar {L} {v^i}\right)\d t +
h_i(v^i\d t-\d q^i)
\eeann
and, for a section $\psi_0$ fulfilling (\ref{psi0}),
\beq
0=\psi _0^*\inn(Y_0)\Omega_0=
\left[
f^i\left(\frac{d p_i}{d t}-\derpar {L} {q^i}\right)+
g^i\left(p_i-\derpar {L} {v^i}\right)+
h_i\left(v^i-\frac{d q^i}{d t}\right)
\right]\d t
\label{cero}
\eeq
reproduces the holonomy condition,
the restricted Legendre map (that is, the definition of the momenta),
and the Euler-Lagrange equations.
The coefficient of the component $f$ vanishes as a consequence of the last equations.

Summarizing, the equation (\ref{psi0}) gives different kinds of
information, depending on the
%type of
verticality of the vector fields $Y_0$ involved.
In particular, we obtain equations of three different~classes:
\ben
  \item
Algebraic (not differential) equations, in coordinates
\(\displaystyle p_i=\derpar {L} {v^i}\), which determine a subset
${\cal W}_1$ of ${\cal W}_0$, where the sections solution must
take their values. These can be called {\sl primary Hamiltonian
constraints}, and in fact they generate, by $\hat\rho_2^0$
projection, the primary constraints of the Hamiltonian formalism
for singular Lagrangians, i.e., the image of the Legendre
transformation, ${\cal F}\Lden( J^1\pi)\subset J^1\pi^*$.
\item
The holonomic differential equations,
in coordinates \(\displaystyle v^i=\frac{d q^i}{d t}\),
forcing the sections solution $\psi_0$ to be lifting of $\pi$-sections. This
property
reflects the fact that the geometric condition
in the unified formalism is stronger  than the usual one in the
Lagrangian formalism.
\item
  The classical Euler-Lagrange equations, in coordinates
\beq
\frac{d}{d t}\left(\derpar {L} {v^i}\right)=
\frac{\partial^2 L }{\partial v^j\partial  v^i}\frac{d^2 q^j}{d t^2}+
\frac{\partial^2 L }{\partial q^j\partial v^i}\frac{d q^j}{d t}+
\frac{\partial^2 L }{\partial t\partial v^i}=
\derpar {L} {q^i}
\label{EL}
\eeq
which are obtained from
\(\displaystyle \frac{d p_i}{d t}=\derpar {L} {q^i}\),
using the previous equations.
\een

\section{The dynamical equations for vector fields}
\protect\label{devf}

\begin{prop}
  The problem of finding sections solutions to (\ref{psi0})
  is equivalent to finding the integral curves of a vector field
  $X_0\in\vf({\cal W}_0)$, which is tangent to ${\cal W}_1$ and satisfies
that
  \beq
  \inn (X_0)\Omega_0=0 \quad ,   \quad
  \inn (X_0)\d t=1 \, .
  \label{mvfuf}
  \eeq
\end{prop}
\proof
In a natural chart in ${\cal W}_0$, the local expression of
a vector field $X_0\in\vf({\cal W}_0)$ is
$$
  X_0
  =f\derpar{}{t}+F^i\derpar{}{q^i}+G^i\derpar{}{v^i}+H_i\derpar{}{p_i} \, .
$$
Then, the second equation (\ref{mvfuf}) leads to $f=1$, and the
first gives
\bea
&\mbox{\rm coefficients in $\d p_i$} :& F^i=v^i
\label{one} \\
&\mbox{\rm coefficients in $\d v^i$} :& p_i=\derpar {L} {v^i}
\label{two} \\
&\mbox{\rm coefficients in $\d q^i$} :& H_i=\derpar {L} {q^i}
\label{three} \\
&\mbox{\rm coefficients in $\d t$} \ :& -F^i\derpar {L}
{q^i}+G^i\left(p_i-\derpar {L} {v^i}\right)+H_iv^i=0 \ .
\label{four}
\eea
Now, if $\psi_0=(t,q^i(t),v^i(t),p_i(t))$ is an integral curve
of $X_0$, we have that \(\displaystyle F^i=\frac{d q^i}{d t}\),
\(\displaystyle G^i=\frac{d v^i}{d t}\), \(\displaystyle H_i=\frac{d
p_i}{d t}\), and then (see equation (\ref{cero})): \bit
\item
Equations (\ref{one}) are the holonomy condition.
\item
The algebraic equations (\ref{two}) are the compatibility conditions
defining ${\cal W}_1$.
\item
Using (\ref{one}) and (\ref{two}), equations (\ref{three})
are the Euler-Lagrange equations (\ref{EL}).
\item
Taking into account (\ref{one}) and (\ref{three}), equation (\ref{four})
holds identically.
\eit
Observe that the condition that $X_0$ (if it exists) must be tangent to
${\cal W}_1$
holds also identically from the above equations, since
$$
0=X_0\left( p_i-\derpar {L} {v^i}\right)=
-\frac{\partial^2 L }{\partial v^i\partial  v^j}G^i-
\frac{\partial^2 L }{\partial t\partial v^j}-
\frac{\partial^2 L }{\partial q^i\partial v^j}v^i
+\derpar {L} {q^j}
\qquad \mbox{\rm (on ${\cal W}_1$)}
$$
are the Euler-Lagrange equations again. Observe that, if $ L $ is a
regular Lagrangian, these~equations allow us to determine the
functions \(\displaystyle G^i=\frac{d v^i}{d t}\). If $ L $ is
singular, then a constraint algorithm must be used in order to
obtain a final constraint submanifold ${\cal W}_f$ (if it exists)
where consistent solutions exist, that is, $X_0$ must be tangent to
${\cal W}_f$ (see \cite{CMC-2002} and Section \ref{semi} for
details). \qed

Now, the equivalence of the unified formalism with the Lagrangian
and Hamiltonian formalisms can be recovered as follows, where
$\vf_{{\cal W}_1}({\cal W}_0)$ is the set of vector fields on ${\cal
W}_0$ with support in ${\cal W}_1$.

\begin{teor}
Let $X_0$ be a vector field in  ${\cal W}_0$ which is the
solution to the equations (\ref{mvfuf}). Then
the vector field $X_\Lden\in\vf( J^1\pi)$, defined by
$X_\Lden\circ\rho_1^0=\Tan\rho_1^0\circ X_0$,
is a holonomic vector field solution to the equations (\ref{ELvf}).

Conversely, every holonomic vector field solution to the equations
(\ref{ELvf}) can be recovered in this way from a vector field
$X_0\in\vf_{{\cal W}_1}({\cal W}_0)$.
\end{teor}
\proof Let $X_0$ be a vector field on ${\cal W}_0$, which is a
solution to (\ref{mvfuf}). As sections $\psi_0\colon\Real\to {\cal
W}_0$ solution to the geometric equation (\ref{psi0}) must take
value in ${\cal W}_1$, then $X_0$ can be identified with a vector
field $X_1\colon{\cal W}_1\to\Tan{\cal W}_1$ (i.e.,
$\Tan\jmath_1\circ X_1=X_0\vert_{{\cal W}_1}$), and hence there
exists $X_\Lden\colon  J^1\pi\to\Tan( J^1\pi)$ such that
$X_1=\Tan(\rho_1^1)^{-1}\circ X_\Lden\in\vf({\cal W}_1)$. Therefore,
as a consequence of the item 1 in Theorem \ref{mainteor1}, for every
section $\psi_0$ solution to (\ref{psi0}), there exists
$X_\Lden^0\in\vf(j^1\phi(\Real))$ such that $\Tan\jmath_{\phi}\circ
X_\Lden^0=X_\Lden\vert_{j^1\phi(\Real)}$, where $\jmath_\phi\colon
j^1\phi(\Real)\to E$ is the natural embedding. So, $X_\Lden$ is
$\bar\pi^1$-transversal and holonomic. Then, bearing in mind that
$\jmath_1^*\Omega_0=\rho_1^{1*}\Omega_\Lden$, we have
$$
\jmath_1^*\inn(X_0)\Omega_0=\inn(X_1)(\jmath_1^*\Omega_0)=
\inn(X_1)(\rho_1^{1*}\Omega_\Lden)=
\rho_1^{1*}\inn(X_\Lden)\Omega_\Lden
$$
then $\inn(X_\Lden)\Omega_\Lden=0$ because $\inn(X_0)\Omega_0=0$.
A similar reasoning leads us to prove that, if
$\inn(X_0)\d t=1$, then $\inn (X_{\Lden})\d t=1$.

Conversely, given a holonomic vector field $X_\Lden$,
from $\inn(X_\Lden)\Omega_{\Lden}=0$, and taking into account the
above chain of equalities, we obtain that
$\inn(X_0)\Omega_0\in [\vf({\cal W}_1)]^0$
(the annihilator of $\vf({\cal W}_1)$).
Moreover, $X_\Lden$ being holonomic, $X_0$ is holonomic,
and then the extra condition
$\inn(Y_0)\inn(X_0)\Omega_0 = 0$ is also fulfilled for every
$Y_0\in\vf^{{\rm V}(\rho^0_1)}({\cal W}_0)$.
Thus, remembering that
$\jmath_1^*\Tan{\cal W}_0=\Tan {\cal W}_1\oplus_{{\cal W}_1}\jmath_1^*{\rm V}(\rho_1^0)$,
we conclude that $\inn(X_0)\Omega_0 = 0$.
To prove that if
$\inn (X_{\Lden})\d t=1$, then $\inn(X_0)\d t=1$ is trivial.
\qed

Finally, the Hamiltonian formalism is recovered
using the second equivalence relations (\ref{equivsol}).
The proof for the almost-regular case
follows in a straightforward way.

\section{An example:  spatial semidiscretization of the nonlinear wave
equation}\label{semi}

Consider the nonlinear wave equation given by
\begin{equation}\label{aqe}
 u_{tt}=\frac{d}{dx}\left(\frac{\partial \sigma}{\partial u_x}(t,u_x)\right)-\frac{\partial g}{\partial u}(t,u)
\end{equation}
where $u: U\subset\R^2\rightarrow \R$, $u(t,x)$ and $\sigma$ and $g$ are smooth
functions and we impose periodic boundary conditions $u(t,x)=u(t, x+K)$, $K>0$. Different choices of the functions $\sigma$ and $g$ idealize one-dimensional models for fluids and materials.

Equation (\ref{aqe}) corresponds to the Euler-Lagrange equation derived  extremizing the action functional
\[
u\mapsto \int_0^T\int_0^K \left(\frac{1}{2}
u_t^2-\sigma(t, u_x)-g(t, u)\right)\, dt\, dx\, ,
\]
where we will assume in the sequel the regularity condition $\displaystyle{ \frac{\partial^2 \sigma}{\partial u_x^2}}\not=0$.

One basic idea towards a geometric discretization \cite{LeRe,LMM-2008,OlWeWu} of this type of equations is first to introduce an spatial truncation, that reduce the PDE (\ref{aqe}) to a system of ODEs preserving many of its geometrical properties.
Hence,  we  replace the $x$-derivative in the Lagrangian by a simple
difference (for simplicity, we will work with a uniform grid of
$N+1$ points, $h=K/N$) as follows:
\[
 L(t, u_i, (u_i)_t)= \sum_{i=0}^{N-1}
\left[\frac{1}{2}\left(\frac{(u_i)_t+(u_{i+1})_t}{2}\right)^2 -\sigma\left(t,
\frac{u_{i+1}-u_i}{h}\right)-g\left(t, \frac{u_{i+1}+u_i}{2}\right)\right],
\]

In a more convenient notation, we are working with  the Lagrangian function $L: \R\times
T\R^{N+1}\longrightarrow \R$:
\[
 L(t, q^i, v^i)= \sum_{i=0}^{N-1}
\left[\frac{1}{2}
\left(v^{i+1/2}\right)^2 -\sigma\left(t,
w^i\right)-g\left(t, q^{i+1/2}\right)\right].
\]
where $w^i=\displaystyle{ \frac{q^{i+1}-q^i}{h}}$  and
$Q^{i+1/2}=\displaystyle{ \frac{Q^{i+1}+Q^i}{2}}$,
$i=0\,\ldots, N-1$, $Q=q,v$. Now, following the notation in
previous sections, we find that
\[
\Theta_0=(L(t, q^i, v^i)-p_iv^i)\, dt +p_i\, dq^i\; , 0\leq
i\leq N.
\]
Consider now a vector field
\[
X_0=f\frac{\partial}{\partial t}+F^i\frac{\partial}{\partial q^i}+ G^i\frac{\partial}{\partial v^i}
+ H_i\frac{\partial}{\partial p_i},
\]
satisfying the equations:
\[
i_{X_0}\Omega_0=0, \qquad i_{X_0}dt=1.
\]
It is easy to deduce that:
\[\left\{
\begin{array}{l}
f=1\\
F^i=v^i,\\
H_0=\displaystyle \frac{1}{h}\frac{\partial \sigma}{\partial u_x}\left(t,
w^0\right)-\frac{1}{2} \frac{\partial g}{\partial u}\left(t, q^{0+1/2}\right)\\
H_i=\displaystyle \frac{1}{h}\left(\frac{\partial \sigma}{\partial u_x}(t,
w^{i})-\frac{\partial \sigma}{\partial u_x}(t,
w^{i-1})\right)
-\frac{1}{2}\left( \frac{\partial g}{\partial u}(t, q^{i+1/2})+\frac{\partial g}{\partial u}(t, q^{i-1/2})\right), \quad 1\leq i \leq N-1\\[10pt]
H_N=\displaystyle -\frac{1}{h}\frac{\partial \sigma}{\partial u_x}\left(t,
w^{N-1}\right)-\frac{1}{2} \frac{\partial g}{\partial u}\left(t, q^{N-1/2}\right)
\end{array}\right.
\]
and the constraints defining ${\mathcal W}_1$:
\[
p_0=\frac{1}{2}v^{0+1/2}, \qquad p_i=\frac{1}{2}(v^{i-1/2}+v^{i/2}), \qquad p_N=\frac{1}{2}v^{N-1/2}.
\]
Since $X_0$ must be tangent to ${\mathcal W}_1$ then we obtain the additional conditions
\begin{eqnarray*}
0&=&X_0(p_0-\frac{1}{2}v^{0+1/2})=-\frac{G^0+G^1}{4}+\frac{1}{h}\frac{\partial
\sigma}{\partial u_x}\left(t,
w^0\right)-\frac{1}{2} \frac{\partial g}{\partial u}\left(t, q^{0+1/2}\right)\\
0&=&X_0(p_i-\frac{1}{2}(v^{i-1/2}+v^{i/2}))=-\frac{G^{i-1}+2G^i+G^{i+1}}{4}+
\frac{1}{h}\left(\frac{\partial \sigma}{\partial u_x}(t,
w^{i})-\frac{\partial \sigma}{\partial u_x}(t,
w^{i-1})\right)
\\
&&-\frac{1}{2}\left( \frac{\partial g}{\partial u}(t, q^{i+1/2})+\frac{\partial g}{\partial u}(t, q^{i-1/2})\right), \quad 1\leq i \leq N-1\\[10pt]
0&=&X_0(p_N-\frac{1}{2}v^{N-1/2})=-\frac{G^{N-1}+G^N}{4}-\frac{1}{h}\frac{\partial
\sigma}{\partial u_x}\left(t, w^{N-1}\right)-\frac{1}{2}
\frac{\partial g}{\partial u}\left(t, q^{N-1/2}\right)
\end{eqnarray*}

From these last equations we obtain $G^0, G^1, \cdots, G^{N-1}$ in terms of $G^N$ and the additional constraint
\[
\sum_{i=0}^{N-1} (-1)^i\frac{\partial \sigma}{\partial u_x}\left(t,
w^{i}\right)=0
\]
which determines the new constraint submanifold, ${\mathcal W}_2$. Again, the condition of tangency of $X_0$ to ${\mathcal W}_2$ gives us a new constraint:
\[
\sum_{i=0}^{N-1}(-1)^i\left[\frac{\partial^2
\sigma}{\partial u_x\partial t}\left(t,
w^{i}\right)+\frac{v^{i+1}-v^i}{h}\frac{\partial^2
\sigma}{\partial u_x^2}\left(t, w^{i}\right)\right]=0.
\]
determining the constraint submanifold, ${\mathcal W}_3$.
 From it, we obtain that
 \begin{eqnarray*}
\displaystyle \sum_{i=0}^{N-1}(-1)^i\left[ \frac{\partial^3
\sigma}{\partial u_x\partial^2 t}\left(t,
w^i\right)\right.&+&\frac{v^{i+1}-v^{i}}{h}\frac{\partial^3
\sigma}{\partial u_x^2\partial t}\left(t,
w^{i}\right)\\
&&\displaystyle\left.+\left(\frac{v^{i+1}-v^{i}}{h}\right)^2\frac{\partial^3
\sigma}{\partial u_x^3}\left(t, w^{i}\right)
+\frac{G^{i+1}-G^{i}}{h}\frac{\partial^2 \sigma}{\partial
u_x^2}\left(t, w^{i}\right)\right]=0
\end{eqnarray*}
which uniquely determines  the remaining coefficient $G^N$ form the regularity condition $\displaystyle{ \frac{\partial^2 \sigma}{\partial u_x^2}}\not=0$.
%Therefore, ${\mathcal W}_3$

\section{Conclusion and outlook}
\protect\label{di}

Following the Skinner-Rusk model for autonomous mechanical systems,
we have presented a generalized framework for describing
both Lagrangian and Hamiltonian time dependent mechanical systems.

The key tool of this construction is the coupling form
which is defined using the natural geometric structure of the
manifold ${\cal W}= J^1\pi \times_{E}\Tan^*E$. This function
allows us to define in a natural way a submanifold
${\cal W}_0$ of ${\cal W}$, which is diffeomorphic to
${\cal W}_r= J^1\pi \times_{E}J^1\pi^*$,
the true space of physical variables.
Then, the compatibility of the dynamical equations stated in ${\cal W}_0$
gives a new submanifold ${\cal W}_1$ which is identified
with the graph of the Legendre map $\widetilde{{\cal F}\Lden}$,
where all the characteristic features of the Lagrangian and Hamiltonian
formalisms of time-dependent regular and singular non-autonomous
systems are recovered.

This unified formalism constitutes an alternative but equivalent approach to
that given by Cantrijn {\it et al} in \cite{CMC-2002}.
The essential difference is that, in this work, the
dynamical equations are established directly in ${\cal W}= J^1\pi \times_{E}\Tan^*E$.
These equations are compatible in a $1$-codimensional submanifold of ${\cal W}$,
but the dynamical solution is undetermined, even in the regular case.
In order to overcome this trouble, the authors are forced
to introduce a new constraint, in such a way that
the resulting submanifold is the graph of the Legendre map.
As a consequence, they are unable to define intrinsically the submanifold
of physical states ${\cal W}_0$.
In our model, the introduction of the coupling form
gets round all the above problems.

The Skinner-Rusk unified formalism which is developed here
has been used to give a new geometric framework for
time-dependent optimal control problems in
\cite{BEMMR-2007}, where some interesting examples are
analyzed. Following the above example the developed
formalism could be applied to optimal control problems in
partial differential equations where the spatial
semidiscretization is used to solve.

\subsection*{Acknowledgments}

We acknowledge the financial support of \emph{Ministerio de
Educaci\'on y Ciencia}, Projects MTM2005-04947,
MTM2007-62478, and  S-0505/ESP/0158 of the CAM. One of us (MBL)
also acknowledges the financial support of the FPU grant
AP20040096. We thank Mr. Jeff Palmer for his assistance in
preparing the English version of the manuscript.


{\small

\begin{thebibliography}{99}

\bibitem{BEMMR-2007}
{\sc M. Barbero-Li\~n\'an, A. Echeverr\'\i a-Enr\'\i quez,
D. Mart\' \i n de Diego, M.C. Mu\~noz-Lecanda, N. rom\'an-Roy},
``Skinner-Rusk unified formalism for optimal control problems and applications'',
{\sl J. Phys. A: Math. Theor.} {\bf 40} (2007) 12071-12093.

\bibitem{CMC-2002}
{\sc  F. Cantrijn, J. Cort\'es, S. Mart\'\i nez}, ``Skinner-Rusk
approach to time-dependent mechanics'', {\sl Phys. Lett. A} {\bf
300} (2002) 250-258.

\bibitem{CLMM-2002}
{\sc J. Cort\'es, M. de Le\'on, D. Mart\'\i n de Diego, S.
Mart\'\i nez}, ``Geometric description of vakonomic and
nonholonomic dynamics. Comparison of solutions''. {\sl SIAM J.
Control and Optimization} (to appear) (2002).

\bibitem{ELMMR-04}
{\sc A. Echeverr\'\i a-Enr\'\i quez, C. L\'opez, J. Mar\'\i
n-Solano, M.C. Mu\~noz-Lecanda, N. Rom\'an-Roy},
``Lagrangian-Hamiltonian unified formalism for field theory'',
{\sl J. Math. Phys.} {\bf 45}(1) (2004) 360-385.

\bibitem{EMR-91}
{\sc A. Echeverr\'\i a-Enr\'\i quez, M.C. Mu\~noz-Lecanda, N.
Rom\'an-Roy}, ``Geometrical setting of time-dependent regular
systems. Alternative models'', {\sl Rev. Math. Phys.} {\bf 3}(3)
(1991) 301-330.

\bibitem{GM-05}
{\sc X. Gr\`acia, R. Mart\'\i n}, ``Geometric aspects of
time-dependent singular differential equations'', {\sl Int. J.
Geom. Methods Mod. Phys.} {\bf 2}(4) (2005) 597-618.

\bibitem{Ku-tdms}
{\sc R. Kuwabara}, ``Time-dependent mechanical symmetries and
extended Hamiltonian systems'', {\sl Rep. Math. Phys.} {\bf 19}
(1984) 27-38.


\bibitem{LeRe} {\sc B. Leimkuhler, S. Reich}: {\sl Simulating Hamiltonian Dynamics}, Cambridge Monographs on Applied and Computational Mathematics, Cambridge University Press, 2004.

\bibitem{LMM-2002}
{\sc M. de Le\'on, J.C. Marrero, D. Mart\'\i n de Diego}, ``A new
geometrical setting for classical field theories'', {\sl
Classical and Quantum Integrability}. Banach Center Pub. {\bf 59},
Inst. of Math., Polish Acad. Sci., Warsawa (2002) 189-209.


\bibitem{LMM-2008} {\sc M. de Le\'on, J.C. Marrero, D. Mart\'\i n de Diego}, ``Some applications of semi-discrete variational integrators to classical field theories", To appear in  Qualitative Theory of Dynamical Systems (2008).

\bibitem{LR} {\sc M. de Le\'on, P.R. Rodrigues},
{\sl  Methods of Differential Geometry in Analytical Mechanics},
North-Holland Math. Ser. 152, Amsterdam, 1989.

\bibitem{MS-98}
{\sc L. Mangiarotti, G. Sardanashvily}, ``Gauge Mechanics'',
{\sl World Scientific}, Singapore, 1998.

\bibitem{OlWeWu} {\sc M. Oliver, M. West, C. Wulff}, ``Approximate momentum conservation for spatial semidiscretizations of semilinear wave equations'',    Numerische Mathematik {\bf 97} (3) (2004), 493-535

\bibitem{Ra1}
{\sc M. F. Ra\~nada}, ``Extended Legendre transformation approach
to the time-dependent Hamiltonian formalism'', {\sl J. Phys. A:
Math. Gen.} {\bf 25} (1992) 4025-4035.

\bibitem{RRS}
{\sc A.M. Rey, N. Rom\'{a}n-Roy, M. Salgado},
``G\"{u}nther's formalism in classical
field theory: Skinner-Rusk approach and the evolution operator'',
{\sl J. Math. Phys.} {\bf 46}(5) (2005) 052901.

\bibitem{St-2005}
{\sc J. Struckmeier}, ``Hamiltonian dynamics on the symplectic
extended phase space for autonomous and non-autonomous systems'',
{\sl J. Phys. A: Math. Gen.} {\bf 38} (2005) 1275--1278.

\bibitem{Sa-89}
{\sc D.J. Saunders}, {\sl The Geometry of Jet Bundles}, London
Math. Soc. Lect. Notes Ser. {\bf 142}, Cambridge, Univ. Press,
1989.

\bibitem{SR-83}
{\sc R. Skinner, R. Rusk}, Generalized Hamiltonian dynamics I:
Formulation on $T^*Q\otimes TQ$'', {\sl J. Math. Phys.} {\bf 24}
(1983) 2589-2594.

%\bibitem{Tulczy1}
%{\sc W.M. Tulczyjew}, ``Hamiltonian systems, Lagrangian systems
%and the Legendre transformation'', {\sl Symposia Mathematica}
%{\bf 16} (1974) 247--258.


\end {thebibliography}

}
\end{document}